\documentclass[preprint,12pt]{elsarticle}

\usepackage{amssymb}
\usepackage{setspace}
\begin{document}

\begin{frontmatter}

%% Title, authors and addresses

%% use the tnoteref command within \title for footnotes;
%% use the tnotetext command for the associated footnote;
%% use the fnref command within \author or \address for footnotes;
%% use the fntext command for the associated footnote;
%% use the corref command within \author for corresponding author footnotes;
%% use the cortext command for the associated footnote;
%% use the ead command for the email address,
%% and the form \ead[url] for the home page:
%%
%% \title{Title\tnoteref{label1}}
%% \tnotetext[label1]{}
%% \author{Name\corref{cor1}\fnref{label2}}

 \ead{michel.planat@femto-st.fr}

%% \ead[url]{home page}
%% \fntext[label2]{}
%% \cortext[cor1]{}
%% \address{Address\fnref{label3}}
%% \fntext[label3]{}

\title{On the geometry and invariants \\ of qubits, quartits and octits}

%% use optional labels to link authors explicitly to addresses:
%% \author[label1,label2]{<author name>}
%% \address[label1]{<address>}
%% \address[label2]{<address>}

\author{M. Planat}

\address{Institut FEMTO-ST/CNRS, 32 Avenue de l'Observatoire, 25044 Besan\c{c}on Cedex, France\\michel.planat@femto-st.fr}

\begin{abstract}
Four level quantum systems, known as quartits, and their relation to two-qubit systems are investigated group theoretically. Following the spirit of Klein's lectures on the icosahedron and their relation to Hopf sphere fibrations, invariants of complex reflection groups occuring in the theory of qubits and quartits are displayed. Then, real gates over octits leading to the Weyl group of $E_8$ and its invariants are derived. Even multilevel systems are of interest in the context of solid state nuclear magnetic resonance.

%Then, the commutation relations of systems involving quartits are shown to be controled by free modules and the projective line over a appropriate ring. Quartits involve the modular ring $\mathbb{Z}_4$, a composite qutrit/quartit system involves the ring $ \mathbb{Z}_3 \times \mathbb{Z}_4$, a qubit/quartit system involves the ring $  \mathbb{Z}_4 \times \mathbb{Z}_4$ and so on (Part II). Quartit systems are of interest in the context of solid state nuclear magnetic resonance.

\end{abstract}

\begin{keyword}
%% keywords here, in the form: keyword \sep keyword
Multilevel systems, quartits, finite groups, invariant theory, $E_8$ Weyl group. 
%% MSC codes here, in the form: \MSC code \sep code
%% or \MSC[2008] code \sep code (2000 is the default)
%\PACS 03.65 Ud \sep 03.67.Pp \sep 02.20.-a

\end{keyword}

\end{frontmatter}

%%
%% Start line numbering here if you want
%%
% \linenumbers

%% main text
\section{Introduction}
\label{}

A four-level system, also denoted a quartit, is a promising concept for the design of versatile two-qubit states and gates of quantum computation. As shown for instance in \cite{Hirayama06}, states of nuclear spin $\frac{3}{2}$
in a specific GaAs quantum well device may be used for realizing the logical single and two-qubit gates by applying selective radio frequency pulses at the resonance frequency between two energy levels. Thus, quantum computing based on solid state nuclear magnetic resonance (NMR) with spins $3/2$ (corresponding to quartits), spins $7/2$ (corresponding to $8$-level systems, or octits) and higher order spins, is a strong motivation for our research. 

Multilevel systems with a number $d$ of levels not a prime number (such as quartits, sextits and so on) are special by the mere fact that the number of mutually unbiased bases (MUBs) that can be constructed from the machinery of finite fields is strictly less than the value $d+1$ corresponding to a complete set. It is well known that a complete set of MUBs may be obtained for a system of $m$ qudits, if the qudit is a $p$-level system, i.e. $d=p^m$ and $p$ a prime number \cite{Planat06}. It has been shown in previous works \cite{Planat08, Havlicek08} that the free modules and the projective line over the finite ring $\mathbb{Z}_2 \times \mathbb{Z}_3 $ may be used to represent the geometry of commutation relations over a sextit system, leading to a maximum of three MUBs. In the context of MUBs, the sextit system is relevant because it corresponds to the smallest composite dimension for which a complete set cannot, in principle, be derived. But the sextit system cannot be ditinguished from the qubit/qutrit system \cite{Havlicek08} so that, from the point of view of quantum computation, it does not add new features.

In contrast, in dimension four, a two-qubit system is not equivalent to a quartit system. A two-qubit system is obtained by taking tensor products of ordinary Pauli spin matrices $\sigma_x$, $\sigma_y$ and $\sigma_z$, while the quartit system is generated by the two {\it shift} and {\it clock} operators (see Example 7 in \cite{Kibler2009})
\begin{equation}
X=\left(\begin{array}{cccc} 0 &0 &0 & 1 \\1 & 0  &0 & 0 \\0 & 1 & 0 & 0 \\0& 0 & 1 & 0\\ \end{array}\right),~~ Z= \mbox{diag}(1,\omega,\omega^2,\omega^3) = \sigma_z \otimes P,
\end{equation}
where $ \omega=\exp(\frac{2i \pi}{4})$ and $P=\mbox{diag}(1,i)$.

The two-qubit Pauli group $\mathcal{P}_2$ is generated by the two-fold tensor products of ordinary Pauli spin matrices. It is isomorphic to the small permutation group $[64,266]$, the group of number $266$ in the 
sequence of small groups with cardinality $64$. It may also be seen as a central product \footnote{In this paper, symbols $ \times$, $*$, $\rtimes$ and $.$ denote the direct, central, semidirect and dot product of groups, respectively \cite{Magma} }: $\mathcal{P}_2 \cong E_{32}^{\pm} * \mathbb{Z}_4$ since $[64,266]$ contain the extraspecial groups $E_{32}^{\pm}$ and the cyclic group $\mathbb{Z}_4$ as normal subgroups, and $E_{32}^{\pm}  \cap \mathbb{Z}_4 $ coincides with the center $\mathbb{Z}_4$.

The quartit group $\mathcal{P}_{quartit}$, generated by matrices $X$ and $Z$, is isomorphic to the small group $[64,18]$. It may also be seen as a semidirect product: $\mathcal{P}_{quartit} \cong \mathbb{Z}_4^2 \rtimes \mathbb{Z}_4$ ( or as a central product $ \mathbb{Z}_4^2 * \mathbb{Z}_4$).

%In group theoretical language, the two-qubit Pauli group $\mathcal{P}_2$ (which is generated by the tensor product of Pauli spin matrices) is isomorphic to the small permutation group $[64,266]$, the group of number $266$ in the 
%sequence of small groups of cardinality $64$. It may also be seen as the semidirect product $\mathcal{P}_2 \cong [16,13] \rtimes \mathbb{Z}_2^2$ of the small group $[16,13]\cong \mathcal{P}_1$ (i.e. isomorphic to the single qubit %Pauli group $\mathcal{P}_1$) by the Klein group $\mathbb{Z}_2^2$. 

The reason why quartit actions relate to two-qubit actions is made explicit in \cite{Hirayama06,Kessel99}. In particular, the standard (entangling) cnot gate is easily inplemented by applying a suitable r.f. $\pi$ pulse between two of the levels of the quartit system (see Sec. 3.3 of \cite {Hirayama06}). This may be reformulated in the group language. The group $\mathcal{C_{\mbox{cnot}}}=\left\langle X,Z,\mbox{cnot}\right\rangle$ obtained from the quartit group by adding to it the cnot generator reads $\mathcal{C}_{\mbox{cnot}} \cong \mathcal{P}_2 \rtimes (\mathbb{Z}_2 \times S_4)$ (of order 3072). This means that two-qubit Pauli operators arise naturally in the normal subgroup part of $\mathcal{C}_{\mbox{not}}$, as they do in the two-qubit Clifford group $\mathcal{C}_2 \cong \mathcal{P}_2 \rtimes (\mathbb{Z}_2 \times S_6)$ (see (14) of \cite{PlanatKibler}). The fourfold symmetry attached to the quartit system is visible  in the factor group $\mathcal{C_{\mbox{cnot}}}/\mathcal{P}_2$, that contains the symmetric group $S_4$; similarly, the corresponding factor group for the two-qubit system contains the symmetric group $S_6$ ( see also \cite{PlanatSaniga} about the sixfold symmetry of the two-qubit system). 

The connection of the quartit system to the two-qubit Clifford group is made more stringent by using (instead of the cnot gate) the entangling gate $S=\left(\begin{array}{cccc} 1 &-1 &1 & 1 \\1 & 1  &-1 & 1 \\1 & -1 & -1 & -1 \\1& 1 & 1 & -1\\ \end{array}\right)$, that we introduced in several contexts  \cite{Planat2010, Planat2010lat}. In some respect, gate $S$ may be seen as a generalization of the single qubit Hadamard gate $H=\frac{1}{\sqrt{2}}\left(\begin{array}{cc} 1 &1 \\1 & -1   \\ \end{array}\right)$. The real gate $S$ encodes the simultaneous (entangled) eigenvectors of the triple $\left\{\sigma_x \otimes \sigma_x,\sigma_y \otimes \sigma_y,\sigma_z \otimes \sigma_z\right\}$ \cite{Planat2010} and lies in the automorphism group of the lattices $\mathbb{Z}^4$ (see [2] in \cite{Planat2010lat}) and $D_4$ (see [7] in \cite{Planat2010lat}).

The single qubit Clifford group is generated by $H$ and $P$ as $\mathcal{C}_1=\left\langle H,P\right\rangle \cong \mathcal{P}_1 \rtimes D_6$ (with $D_6$ the $12$-element dihedral group) identifies to the rank $2$ complex reflection group $U_9$ (number $9$ in the Shephard-Todd sequence \cite{PlanatKibler}). Similarly, it is straigthforward to check that the group $\mathcal{C_{\mbox{S}}}=\left\langle X,Z,S\right\rangle \cong \mathcal{P}_2 \rtimes S_6$, obtained by completing the quartit group with S (instead of cnot) is a subgroup of index two in $\mathcal{C}_2$, identifies  to the rank $4$ complex reflection group $U_{31}$ (number $31$ in the Shephard-Todd sequence, see also p. 8, footnote g  in \cite{PlanatKibler}). 

\section{Invariants involved in the theory of a single qubit}

There is a well known connection of invariant theory to Clifford groups and self-dual codes \cite{Nebe2001,Nebe2006}. In the code context, the genus-$n$ full weight enumerators of some self dual codes are polynomials in $2^n$ variables invariant under the Clifford group $\mathcal{C}_n$. Here, we are interested in a explicit connection of polynomial invariants of $\mathcal{C}_1$ and $\mathcal{C}_S$ to quantum states of the single qubit and quartits respectively. In this goal, it is useful to recollect the celebrated Klein's lectures about the homogeneous invariants attached to the geometry of platonic solids \cite{Klein1956} and give them an interpretation based on the Bloch sphere \cite{Mosseri2001}.  Among Klein's invariants, fundamental invariants of $\mathcal{C}_1$ (of degrees $8$ and $24$)  are polynomials in the two amplitudes of a single qubit state, as shown  in Sec. \ref{singlequbit}. The reasoning is  generalized, in Sec. \ref{singlequartit}, to the fundamental invariants of the complex reflection group $\mathcal{C}_S$ (of degrees $8$, $12$, $20$ and $24$), which are polynomials in the four amplitudes of the quartit.

\subsection{Invariant ring of a finite matrix group}
\label{invring}
Beforehand, let us recall some technicalities. An algorithm that calculates invariant rings of finite linear groups over an arbitrary field $\mathbb{K}$ is described in \cite{Kemper1996} and implemented in Magma \cite{Magma} (see also \cite{Nebe2006,Kane2001,King2007}).

It may be applied to any finite matrix group $G\le GL_n(\mathbb{K})$ of degree $n$ over $\mathbb{K}$. The group $G$ acts linearly on the polynomial ring $\mathbb{K}[x_1,\ldots,x_n]$ of the variables $x_i$ \footnote{In the sequel of this paper, variables $x_i$ are interpreted as the complex amplitudes of a qudit and are denoted $\alpha$, $\beta$ (for a qubit) and $\alpha$, $\beta$, $\gamma$, $\delta$ (for a quartit).}.  The invariant ring is defined as the set of polynomials left invariant under the action of $G$
\begin{equation}
R=\mathbb{K}[x_1,\ldots,x_n]^G:=\left\{f \in \mathbb{K}[x_1,\ldots,x_n]|\sigma(f)=f,~ \forall \sigma \in G\right\}.
\end{equation}
Primary invariants may be constructed, i.e., homogeneous invariants $f_1,\ldots,f_n$ that are algebraically independant, such that the invariant ring is a finitely generated homogeneous module over $A=\mathbb{K}[f_1,\ldots,f_n]$.

Then, a (minimal) set of secondary invariants for $R$, with respect to these primary invariants, is a (minimal) generating set for $R$ considered as a module over the algebra generated by the primary invariants. 

Given $I$ a homogeneous ideal of the graded polynomial ring $P=\mathbb{K}[x_1,\ldots,x_n]$, then the quotient ring $P/I$ is a graded vector space: $P/I$ is a direct sum of the vector spaces $V_d$ for $d=0,1,\ldots$ where $V_d$ is the $\mathbb{K}$-vector space consisting of all homogeneous polynomials in $P/I$ of weighted degree d. The Hilbert series of the graded vector space $P/I$ is the generating fonction
$$H_{P/I}(t)=\sum_{d=0}^{\infty}\mbox{dim}(V_d)t^d,$$
and may be rewritten as a rational function of the variable $t$.

If the character $\mbox{char}(\mathbb{K})$ of $\mathbb{K}$ does not divide the cardinality $|G|$ of the group $G$ (the non-modular case), the Hilbert series becomes the Molien series \cite{Kemper1996}
$$H_t=\frac{1}{|G|}\sum_{\sigma \in G}\frac{1}{\mbox{det}(1-\sigma t)}.$$

\subsection{Geometry of the invariants of the single qubit Clifford group $\mathcal{C}_1$}
\label{singlequbit}

Let us see how the same results follow from the geometrical approach \cite{Klein1956}. In the computational basis, a single qubit reads as the superposition
\begin{equation}
\left| \psi\ \right\rangle=\alpha \left|0\ \right\rangle + \beta\left|1\ \right\rangle,~~\alpha,\beta \in \mathbb{C},~~|\alpha|^2+|\beta|^2=1,
\label{qubits}
\end{equation}
where the complex amplitudes $\alpha=a+ib$, $\beta=c+id$ satisfy the equations of the $3$-sphere $S^3: a^2+b^2+c^2+d^2=1$. This state may also be represented (up to a global phase) in the Bloch sphere picture by using the map \cite{Mosseri2001}
\begin{eqnarray}
&\xi=\left\langle \sigma_x\right\rangle_{\psi}=2~\mbox{Re}(\bar{\alpha}\beta),\nonumber\\
&\eta=\left\langle \sigma_y \right\rangle_{\psi}=2~\mbox{Im}(\bar{\alpha}\beta),\nonumber\\
&\zeta=\left\langle \sigma_z \right\rangle_{\psi}=|\alpha|^2-|\beta|^2, \nonumber\\
\end{eqnarray}
satisfying the equation of the $2$-sphere $S^2:\xi^2+\eta^2+\zeta^2=1$.
Specific points are the north pole $\left|0\ \right\rangle$, the south pole  $\left|1\ \right\rangle$, and a set $\mathcal{E}$ of four points of the equatorial plane located at the intersection of the Bloch sphere with the reference axes $(\pm 1,0,0)=\frac{1}{\sqrt{2}}(\left|0\ \right\rangle\pm\left|1\ \right\rangle)$,  $(0,\pm 1,0)=\frac{1}{\sqrt{2}}(i\left|0\ \right\rangle \pm\left|1\ \right\rangle)$.

The Bloch sphere picture defines the Hopf fibration $S^3 \stackrel{S^1} {\rightarrow} S^2$, in which the (great circle) fibre $S^1$ represents the global phase.

%Let us start from the representation of a single qubit in the Bloch sphere picture. The generic qubit state is written as a linear superposition of basis %states $\left|0\ \right\rangle$ and $\left|1\ \right\rangle$ 
%
%\begin{equation}
%\left| \psi\ \right\rangle=\cos \frac{\theta}{2}\left|0\ \right\rangle + e^{i \phi}\sin \frac{\theta}{2}\left|1\ \right\rangle,
%\end{equation}
%
%parametrized by the continuous variables $0\le\theta\le\pi$ and $0\le\phi < 2\pi$.
%In the Cartesian coordinates $(\xi=\cos\phi \sin \theta,\eta=\sin \phi\sin \theta,\zeta=\cos \theta)$, the qubit may be seen as an inhabitant of the %$2$-sphere of equation $\xi^2+\eta^2+\zeta^2=1$, called the Bloch sphere.

The picture of the Bloch sphere may be supplemented by the picture of the Riemann sphere $\mathbb{C}\cup \{\infty \}$%or, more abstractly, by the picture of the complex projective line $\mathbb{C}\mathbb{P}^1$. Every rotation of the sphere corresponds to an element of the continuous group $\mbox{SU}(2)$, i.e. to a fractional linear transformation (in two complex variables $\alpha$ and $\beta$), acting on a generic point $x+i y$ of the equatorial plane taken in complex coordinates $x$ and $y$ as
\begin{equation}
\left(\begin{array}{c} x' \\y' \\  \end{array}\right)=\left(\begin{array}{cc} \alpha & \beta \\ -\bar{\beta} & \bar{\alpha}  \\ \end{array}\right)
\left(\begin{array}{c} x \\y \\  \end{array}\right),~~\left|\alpha \right|^2+\left|\beta \right|^2=1.
\end{equation}
Then, following Klein \cite{Klein1956}, points of the Riemann sphere minus the north pole $(0,0,1)$ are mapped to the equatorial plane $\zeta=0$, via a stereographic projection 
\begin{equation}
s(\xi,\eta,\zeta)=\frac{\xi+i\eta}{1-\zeta},
\label{stereo}
\end{equation}
and the north pole is mapped to the {\it fraction} $s(0,0,1)=\frac{1}{0}.$

In this representation, a point of the sphere $S^2$ can be seen as a equivalence class $\frac{p}{q}$ (so that to fractions $\frac{p}{q}$ and $\frac{p'}{q'}$, such that $pq'-p'q=1$, represent the same point on the sphere). For this reason, the Riemann sphere may also be identified to the complex projective line $\mathbb{C}\mathbb{P}^1$.  

As a result, the {\it fraction} $\frac{p}{q}$ of the Riemann sphere picture corresponds to the qubit $\alpha \left|0\ \right\rangle+\beta \left|1\ \right\rangle$ of the Bloch sphere picture. The specific points above are the north pole $\frac{1}{0}\equiv \left|0\ \right\rangle$, the south pole  $\frac{0}{1}\equiv \left|1\ \right\rangle$, and points of the set $\mathcal{E}$ that are
 $s(\pm 1,0,0)=\pm\frac{1}{1}\equiv
 \frac{1}{\sqrt{2}}(\left|0\ \right\rangle\pm\left|1\ \right\rangle)
 $ 
and $s(0,\pm 1 ,0)=\pm \frac{i}{ 1}\equiv\frac{1}{\sqrt{2}}(\left|0\ \right\rangle\pm i\left|1\ \right\rangle)$.

Having parametrized the $2$-sphere as $\mathbb{C}\mathbb{P}^1$, one can capture the symmetry of a set of $n$ points $\mathcal{S}=\left\{\frac{p_1}{q_1},\cdots,\frac{p_n}{q_n}\right\}$, living on it, in a polynomial of the form \cite{Klein1956,Weeks2007} 
\begin{equation}
(\frac{\alpha}{\beta}-\frac{p_1}{q_1})\cdots (\frac{\alpha}{\beta}-\frac{p_n}{q_n}),
\label{Klein}
\end{equation}
that has roots exactly at the points of $\mathcal{S}$. This can be rewritten as a homogeneous polynomial of degree $n$
\begin{equation}
(q_1\alpha-p_1\beta) \cdots (q_n\alpha-p_n \beta). 
\label{poly}
\end{equation}
The pairs $(\alpha,\beta)$ which cancel the polynomial (\ref{poly}) are the amplitudes of the quantum states in (\ref{qubits}). 
%One also needs to identify the discrete subgroup of $\mbox{SU}(2)$ leaving invariant this polynomial.
For instance, the set of six points listed above may be seen as the vertices of a octahedron and their symmetry may be expressed as the polynomial
\begin{equation}
\mathcal{T}:=\alpha\beta(\alpha-\beta)(\alpha+\beta)(\alpha+i\beta)(\alpha-i\beta)=\alpha\beta(\beta^4-\alpha^4),
\label{inv1}
\end{equation}
which is invariant under the action of the octahedral group. 

\subsubsection*{The tetrahedral group} 

Refering to the specific points displayed at the previous section, the faces of the octahedron are centered at the vertices of a cube of coordinates
\begin{equation}
\pm \xi=\pm \eta=\pm \zeta=\frac{1}{\sqrt{3}},
\label{faces}
\end{equation}
and one can pick up the vertices of a tetrahedron by selecting, among the eight possible combinations of sign, those for which the product $\xi\eta\zeta$ is positive
\begin{equation}
\frac{p}{q}=\frac{1+i}{\sqrt{3}-1},~\frac{1-i}{\sqrt{3}+1},\frac{-1+i}{\sqrt{3}+1},\frac{-1-i}{\sqrt{3}-1}.
\label{fourvertices}
\end{equation}

 A straightforward calculation leads to the invariant polynomials
\begin{equation}
\alpha^4\pm 2i\sqrt{3}\alpha^2\beta^2+\beta^4,
\label{inv2}
\end{equation}
which correspond to a tetrahedron (upper sign) or to a countertetrahedron (lower sign) (see \cite{Klein1956}, p 54).  
One can check that the tetrahedral group $U_4 \cong \mbox{SL}(2,3)$ (number $4$ in the Shephard-Todd sequence)
\small
\begin{equation}
U_4=\left\langle \frac{1}{2}\left(\begin{array}{cc} 1-i & 1-i \\ -\sqrt{3}-1 & 1+\sqrt{3}  \\ \end{array}\right),
\frac{1}{2}\left(\begin{array}{cc} 1-i&i-1\\ 1+\sqrt{3} &1+\sqrt{3}\\ \end{array}\right)
\right\rangle
\end{equation}
\normalsize
possesses an invariant ring spanned by the two primary invariants (\ref{inv1}) and (\ref{inv2}), corresponding to the Molien Series
\begin{equation}
\mbox{MS}(U_4)=\frac{1}{(1-t^4)(1-t^6)}.
\end{equation}

The tetrahedral system is involved in the construction of minimal four-state quantum tomography. The quantum states correspond to the four vertices of the tetrahedron and form a SIC POVM (see eq (2.6) in \cite{Reha04} and Sec. B in \cite{Durt08}). They may be taken as
\begin{eqnarray}
&\left| \psi\ \right\rangle=\alpha \left|0\ \right\rangle + \beta\left|1\ \right\rangle,~\sigma_x\left| \psi\ \right\rangle,~\sigma_y \left| \psi\ \right\rangle,~\sigma_z \left| \psi\ \right\rangle, \nonumber \\
&\mbox{with}~\alpha=\frac{1}{\sqrt{2}}(1+1/\sqrt{3})^{1/2}~\mbox{and}~\beta=\frac{1}{\sqrt{2}}\exp (i\pi/4 )(1-1/\sqrt{3})^{1/2},\nonumber \\
\end{eqnarray}
in agreement with the upper equation (\ref{inv2}).

% The tetrahedral group $U_4\cong SL(2,3)$ is a representation of the smallest known  two-dimensional unitary 2-design \cite{Gross07}.       

\subsubsection*{The octahedral group}

The octahedral group is not of the reflection type but is isomorphic to $U_4$ \cite{Bannai99}. It possesses the invariant polynomial (\ref{inv1}) for the vertices and another invariant polynomial related to the $8$ faces of the octahedron centered at the points given in (\ref{faces}). It is obtained by multiplying together the invariants (\ref{inv2}) so that
\begin{equation}
\mathcal{W}:=\alpha^8+14\alpha^4\beta^4+\beta^8.
\label{inv3}
\end{equation}
The octahedral group $\mathcal{O}$ is generated as the derived subgroup $\mathcal{C}_1'$ of the single qubit Clifford group 
\small
\begin{equation}
\mathcal{O}=\left\langle i\sigma_z,\frac{1}{2}\left(\begin{array}{cc} 1-i & i-1 \\ 1+i & 1+i  \\ \end{array}\right)\right\rangle,
\label{octah}
\end{equation}
\normalsize
The Molien series is 
\begin{equation}
\mbox{MS}(\mathcal{O})=\frac{1-t^4+t^8}{(1-t^6)(1-t^8)}.
\end{equation}
and the invariant ring is spanned by (\ref{inv1}) and (\ref{inv3}).

\subsubsection*{The group $U_8$ and the Clifford group $\mathcal{C}_1$}

The reflection group $U_8 \cong \mathcal{Z}_4.S_4$ is a subgroup of index two in the Clifford group $\mathcal{C}_1$

\begin{equation}
U_8=\left\langle P,\frac{1}{2}\left(\begin{array}{cc} 1-i & i-1 \\ 1+i & 1+i  \\ \end{array}\right)\right\rangle,
\label{U8group}
\end{equation}
Its Molien series is 
\begin{equation}
\mbox{MS}(U_8)=\frac{1}{(1-t^8)(1-t^{12})},
\end{equation}
and the invariant ring is spanned by the invariant $\mathcal{W}$ and an invariant of degree $12$
\begin{equation}
\mathcal{\kappa}:=\alpha^{12}-33~\alpha^8\beta^4-33~\beta^4\alpha^8+\beta^{12}.
\label{inv4}
\end{equation}
There exists the following relationship (eq (53) in \cite{Klein1956}) \footnote{Primary invariants $\mathcal{T}$ and $\mathcal{W}$, the secondary invariant $\kappa$ and relation (\ref{rel}) follow from Magma with the following code \lq\lq R:=InvariantRing(U8); PrimaryInvariants(R); SecondaryInvariants(R); Algebra(R); Relations(R);"}
\begin{equation}
108 ~\mathcal{T}^4-\mathcal{W}^3+\mathcal{\kappa}^2=0.
\label{rel}
\end{equation}
Finally, the Clifford group $\mathcal{C}_1$ is the reflection group $U_9$. The Molien series $1/(1-t^8)(1-t^{24})$, is spanned by the three invariants $\mathcal{T}$, $\mathcal{W}$ and $\mathcal{\kappa}$.

In \cite{Bannai99}, the invariant ring $\mathbb{C}[\alpha,\beta]^{\mathcal{C}_1}$ of $\mathcal{C}_1$ is found to be isomorphic to the polynomial ring $\mathbb{C}[E_4,\Delta_{12}]$ generated by the Eisenstein series $E_4$ of weight $4$ and the cusp form $\Delta_{12}$ of weight $12$, but no reference to Klein's work is pointed out. The invariant ring of $\mathcal{C}_1$ is spanned by the weight enumerator $\mathcal{W}$ of the Hamming code $e_8$ and the weight enumerator $\mathcal{G}$ of the Golay code $G_{24}$ \cite{Nebe2001} 
\begin{equation}
\mathcal{G}:=\alpha^{24}+759~\alpha^{16}\beta^{8}+2576~\alpha^{12}\beta^{12}+759~\alpha^8\beta^{16}+\beta^{24},
\label{inv5}
\end{equation}
where $\mathcal{T}^4= \frac{\mathcal{W}^3-\mathcal{G}}{42}$.  
%In the notation of \cite{Nebe2001} the invariant ring of $\mathcal{C}_1$ is
%
%\begin{equation}
%\mbox{Inv}_{\mathcal{C}_1}:=\frac{1}{\mathcal{W},\mathcal{G}}.
%\end{equation}
% 

\section{Invariants involved in the theory of a single quartit} 
\label{singlequartit}
 
Let us define a quartit state as
\begin{equation}
\left| \psi\ \right\rangle=\alpha \left|0\ \right\rangle + \beta\left|1\ \right\rangle +\gamma \left|2\ \right\rangle+\delta \left|3\ \right\rangle,
\label{quartits}
\end{equation}
with $\alpha,\beta,\gamma,\delta\in \mathbb{C}$ and the sum of squared amplitudes equal to unity. 

Formally, one can map a quartit state to a two-qubit state as $\left|0\ \right\rangle \equiv \left|00\ \right\rangle$, $\left|1\ \right\rangle \equiv \left|01\ \right\rangle$, $ \left|2\ \right\rangle \equiv \left|10\ \right\rangle$ and $\left|3\ \right\rangle \equiv \left|11\ \right\rangle$, so that

\begin{equation}
\left| \psi\ \right\rangle=\alpha \left|00\ \right\rangle + \beta\left|01\ \right\rangle+ \gamma\left|10\ \right\rangle+ \delta\left|11\ \right\rangle,
\label{twoQB}
\end{equation}
with $\alpha,\beta,\gamma,\delta\in \mathbb{C}$ and the sum of squared amplitudes equal to unity.
This key identification is carried out in \cite{Hirayama06}. Although a quartit is not, strictly speaking, equivalent to two qubits, both systems are close to each other as soon as the suitable gate action is performed (as shown in the introduction). Experimentally, it is much convenient to have at one's disposal a four-level system (such as the NMR spin $\frac{3}{2}$ state) than a two-qubit system. In the present context of the second Hopf fibration, it is true that a quartit is formally similar to a two-qubit system, and the resulting polynomial invariants also are the same.

 The quantum states are now the inhabitants of the $7$-sphere $S^7$ of unit radius. The latter can be mapped to the $4$-sphere $S^4$ by using ,instead of complex numbers, the quaternions
\begin{equation}
Q_1=\alpha+\beta j,~Q_2=\gamma+\delta j,~\mbox{with}~Q_1,Q_2\in \mathbb{H},
\end{equation}
where $\mathbb{H}$ denotes the (non-commutative) field of quaternions. The multiplication rules are $i^2=j^2=k^2=-1$, $ij=k$, $jk=i$, $ki=j$, and $ji=-k$, $kj=-i$ and $ik=-j$. The conjugate of a quaternion $Q_1$ is $\bar{Q}_1=\alpha- \beta j$ and the norm is $|Q_1|^2=Q_1 \bar{Q}_1$.

In the second Hopf fibration $S^7 \stackrel{S^3} {\rightarrow} S^4$, the coordinates of a point of the target space are $(\xi,\eta,u,v,\zeta)=\left\langle \sigma_i\right\rangle_{\psi}(i=1..5)$ where 
\footnotesize
\begin{equation}
\sigma_1=\sigma_x=\left(\begin{array}{cc} 0 & 1 \\ 1 & 0  \\ \end{array}\right),~~\sigma_{2,3,4}=\left(\begin{array}{cc} 0 & i,j,k \\ -(i,j,k) & 0  \\ \end{array}\right),~~
\sigma_5=\sigma_z=\left(\begin{array}{cc} 1 & 0 \\ 0 & -1  \\ \end{array}\right),
\end{equation}
\normalsize
are {\it quaternion} Pauli matrices \cite{Mosseri2001}. Explicitely
\begin{eqnarray}
&\xi=2~\mbox{Re}(\bar{\alpha}\gamma+\bar{\beta}\delta),~~~\eta=2~\mbox{Im}(\bar{\alpha}\gamma+\bar{\beta}\delta),\nonumber\\ 
&u=2~\mbox{Re}(\alpha\delta-\beta\gamma),~~~v=2~\mbox{Im}(\alpha\delta-\beta\gamma),\nonumber\\ 
&\zeta=|Q_1^2|-|Q_2^2|=\alpha^2-\beta^2+\gamma^2-\delta^2.\nonumber \\
\end{eqnarray}
The notable point is the sensitivity of the second Hopf map to entanglement. Separable states satisfy the condition $\alpha\delta=\beta\gamma$ and therefore are mapped onto a subset $u=v=0$ of pure complex numbers in $\mathbb{H}$. Conversely, the maximally entangled states correspond to a subset of pure quaternions $\xi=\eta=\zeta=0$ in $\mathbb{H}$.

It is known that the projective Hilbert space of two non-entangled qubits is the product of two 2-dimensional spheres $S^2$, each sphere being the Bloch sphere attached to a specific qubit. In terms of the second Hopf fibration, one sees that the space $S^4$ maps to a unit sphere $S^2$ (such that $u=v=0$) which is the Bloch sphere for the first qubit. The second Bloch sphere is recovered from the fibre. The projective Hilbert space for maximally entangled states is known to be $S^3/\mathbb{Z}_2$, i.e. a $3$ sphere such that two opposite points are identified. This has a counterpart in the Bloch sphere picture in the fact that opposite points on $S^3$ corresponds to the same maximally entangled state \cite{Mosseri2001}.  

The Riemann sphere $\mathbb{C}\cup \infty$ may be generalized to the one-point compactification $\mathbb{H}\cup \infty$ of the quaternion numbers, which may be identified to the projective line $\mathbb{H}\mathbb{P}^1\cong S^4$ over the quaternions. To proceed further, one can map points of $\mathbb{H}\mathbb{P}^1$ to the plane $\zeta=0$, via a 4-dimensional stereographic projection 
\begin{equation}
s(\xi,\eta,u,v,\zeta)=\frac{\xi+i\eta+j u+ k v}{1-\zeta}~~\mbox{with}~~s(0,0,0,0,1)=\frac{1}{0}.
\end{equation}
but the generalization of Klein's approach is not straigthforward since a homogeneous polynomial corresponding to (\ref{Klein}) should display four variables $\alpha$, $\beta$, $\gamma$ and $\delta$ instead of two. Fortunately, the  invariant theory of finite linear groups summarized in Sec. \ref{invring} may be applied and results made explicit by using Magma \cite{Magma}. 

The invariant ring of complex reflection group $\mathcal{C}_S \equiv U_{31}$ attached to a quartit is spanned by invariants of degrees $8$, $12$, $20$ and $24$ in the four variables $x_1=\alpha$, $x_2=\beta$, $x_3=\gamma$ and $x_4=\delta$.
The smallest degree invariant is 
\begin{equation}
\mbox{inv}_8:=\Sigma_8+14 ~\Sigma_{4,4}+168~ \Sigma_{2,2,2,2},
\label{inv8}
\end{equation}
in the notations of \cite{Nebe2001,Nebe2006}, i.e. $\Sigma_8=\sum_{i=1}^4 x_i^8$, $\Sigma_{4,4}=\sum_{j>i} x_i^4 x_j^4$ and $\Sigma_{2,2,2,2}=\prod_{i=1}^4 x_i^2$.

Invariant $\mbox{inv}_8$ also represents the genus-two complete weight enumerator of the code $e_8\otimes \mathbb{F}_4$ \cite{Nebe2001}, and indeed generalizes Klein's invariant (\ref{inv3}).

With the same type of notations, higher order invariants are as follows
\begin{eqnarray}
&\mbox{inv}_{12}:=\Sigma_{12} -33~\Sigma_{8,4}+330~\Sigma_{4,4,4}+792~\Sigma_{2,2,2,6}, \nonumber \\
&\mbox{inv}_{20}:=\Sigma_{20}-19~\Sigma_{16,4}-494~\Sigma_{12,8}+380~\Sigma_{12,4,4}+7296~\Sigma_{10,6,2,2}\nonumber \\
&+1710~\Sigma_{8,8,4}+133380~\Sigma_{8,4,4,4}+102144~\Sigma_{6,6,6,2}, \nonumber \\
&\mbox{inv}_{24}:=\Sigma_{24}+759~\Sigma_{16,8}+2576~\Sigma_{12,12}+212520~\Sigma_{12,4,4,4}+340032~\Sigma_{10,6,6,2}\nonumber \\
&+22770~\Sigma_{8,8,8}+1275120~\Sigma_{8,8,4,4}+4080384~\Sigma_{6,6,6,6}\nonumber .\\
\label{invs}
\end{eqnarray}
Note that $\mbox{inv}_{12}$ and $\mbox{inv}_{24}$ generalize $\kappa$ in (\ref{inv4}) and $\mathcal{G}$ in (\ref{inv5}), and that $\mbox{inv}_{24}$ is the genus-$2$ Hamming weight enumerator of the Golay code $G_{24}$.

\section{On the Weyl group of $E_8$ generated from octits}

A relevant example of an eight-level (spin $\frac{7}{2})$ system, here also denoted a octit, is the $^{133}\mbox{Cs}$ in an anisotropic environment \cite{Khitrin2001}. Real gates over a octit are especially interesting because they allow to generate the largest finite complex reflection group $U_{36}\cong W(E_8)$, of order $696729600$,  where $W(E_8)$ denotes the Weyl group of Lie algebra $E_8$ (see also \cite{Planat2010,Planat2010lat} for a similar approach of $E_8$). One obtains
\begin{equation}
U_{36}=\left\langle X_8, I\otimes I\otimes\sigma_z,S_3\right\rangle,
\end{equation}
where $I$ is the $2 \times 2$ identity matrix, $X_8$ is the $8 \times 8$ shift matrix and $S_3$ is a generalization of matrix $S$, that encodes the eigenstates of the  triple of observables $\sigma_z \otimes\left\{ \sigma_x \otimes \sigma_x, \sigma_y \otimes \sigma_y, \sigma_z \otimes \sigma_z\right\}$
\footnotesize
\begin{equation}
S_3=\frac{1}{2}\left(\begin{array}{cccccccc} 0&0&0&0 &1&1&1&-1 \\1&1&1&-1 &0&0&0&0\\0&0&0&0 &1&1&-1&1\\1&-1&1&1 &0&0&0&0\\
1&1&-1&1 &0&0&0&0 \\-1&1&1&1 &0&0&0&0\\0&0&0&0 &1&-1&1&1\\0&0&0&0 &-1&1&1&1\\ \end{array}\right).
\end{equation}
\label{gateS3}
\normalsize
The Molien series for $U_{36}$ reads
\begin{equation}
\prod_{m=2,8,12,14,18,20,24,30}\frac{1}{1-t^m},
\end{equation}
leading to invariants of degrees given by the indices in the product.
The invariants of smallest degree are found as
\begin{eqnarray}
&I_2=\Sigma_2, ~~I_8=\Sigma_8+56~\Sigma_{4,2,2}-42~\Sigma_{4,4}-168~\Sigma_{2,2,2,2}. \nonumber \\
\end{eqnarray}
Higher order invariants could not be obtained with this method on MAGMA, due to a lack of memory, even on a $96$ MB segment of the cluster at our university. For recent work on this subject, see \cite{Talamini2010}.

\section{Conclusion}

Complex reflection groups of the Shephard-Todd sequence happen to be the natural players in the context of the multilevel approach of quantum computation. Although their construction is well documented in mathematics and the coding theory of self dual codes, due to the identification of the invariants to the complete weight enumerators, their introduction for approaching the design of qubits, quartits and octits seems to be novel. The largest complex reflection group $W(E_8)$ naturally appears from the relevant real gates acting over a octit. It is fascinating that fingerprints of the radii of the Gosset cirles over $E_8$ already appeared in an experimental solid state system (Ising chain) \cite{Coldea2010,Kostant2010}. The Gosset polytope of the $E_8$ lattice may be embedded in the Hopf sphere $S^7$ of a two-qubit system, as shown in \cite{Mosseri2007}. The organic relationship of even Euclidean lattices to entanglement is also explored in \cite{Planat2010lat}.

\bibliographystyle{model1a-num-names}
\bibliography{<your-bib-database>}

%% Authors are advised to submit their bibtex database files. They are
%% requested to list a bibtex style file in the manuscript if they do
%% not want to use model1a-num-names.bst.

%% References without bibTeX database:

\end{document}